# A self-compliant selector-free memory with functional bismuth ferrite

Suman Roy[1,2], Subhabrata Das[3,†], Priyanka Sahu[1,4,†], Majid Ahmadi[5,6], Sameer Kumar Mallik[7] Kaushik Ghosh[3], Saroj Prasad Dash[7], Satyaprakash Sahoo[1,2,*]

[1]*Laboratory for Low Dimensional Materials, Institute of Physics, Bhubaneswar-751005, India.*
[2]*Homi Bhabha National Institute, Training School Complex, Anushakti Nagar, Mumbai-400094, India.*
[3]*Quantum Materials & Devices Unit, Institute of Nano Science and Technology, Knowledge City-Sector 81, Mohali 140306, India.*
[4]*School of Physics, Sambalpur University, Jyoti Vihar, Burla, Odisha 768019, India.*
[5]*Zernike Institute for Advanced Materials, University of Groningen, Groningen, The Netherlands.*
[6]*CogniGron center, University of Groningen, Groningen, The Netherlands.*
[7]*Department of Microtechnology and Nanoscience, Chalmers University of Technology, Göteborg SE-41296, Sweden.*

**Abstract**

In-memory computing requires selector-free operation, low variability, and stable multilevel switching, all of which remain challenging to achieve in a single system. Here, we demonstrate a robust selector-free memory based on a multiferroic $BiFeO_3$ that combines self-compliance with stable bipolar switching, high yield, long endurance, and extended retention. We observe that $BiFeO_3$ memristive crossbar arrays become self-compliant through a naturally formed $TiO_x$ interfacial layer, enabling forming-free bipolar switching without external selector devices, addressing major bottlenecks that have limited the practical deployment of resistive memories. The array delivers stable bipolar switching with a 97% device yield, endurance beyond $10^5$ cycles, and data retention exceeding $10^5$ s with low cycle-to-cycle and device-to-device variability. In addition, the devices exhibit 16 well-separated conductance states, enabling 4-bit multilevel operation via precise pulse-controlled programming for digit visualization in edge applications. Noise measurements reveal predominantly 1/f-type behavior with minimal random telegraph noise, indicating robust conductance stability across the programmable window. These results establish $BiFeO_3$ as a practical materials platform for multibit, selector-free crossbar memories offering high reliability, making it well-suited for scalable computing architectures based on functional materials.

***Corresponding author:** sahoo@iopb.res.in*

[†]These two authors contributed equally to this work

## Introduction

The growth of the Internet of Things, big data analysis, and artificial intelligence capabilities has led to a rapid increase in the requirement of massive data storage and processing[1–3]. The demand for in-memory computing using non-volatile memory (NVM) technologies that offer high speed, low power consumption and high storage density has stimulated extensive research into new functional materials and device physics[4–6]. Emerging NVMs, including phase-change random-access memory (PCRAM)[7], magnetic random-access memory (MRAM)[8], ferroelectric random-access memory (FeRAM)[9], and resistive random-access memory (RRAM)[10], have been intensively studied for data storage and computing to address the limitations imposed by the memory wall in von Neumann architectures. However, each of these technologies presents its own limitations. PCRAM requires high reset currents to amorphize the phase-change material, while MRAM requires large write currents, resulting in increased energy consumption and a need to be immune to magnetic fields [11,12]. FeRAM has gained renewed interest with the development of hafnia-based ferroelectric thin films but it suffers from destructive readout, the wake-up effect, and relatively limited endurance[1314]. On the other hand, redox-based RRAM requires forming voltage and exhibits joule heating that leads to the random fluctuations, write/ read variations and drift of conductance due to the stochastic nature of filament formation[15]. However, RRAM-based NVMs standout for their simple structure, high speed, small cell size ($4F^2$), multibit capability characteristics and compatibility with complementary metal-oxide semiconductor (CMOS) technology[16–18].

To integrate RRAMs into CMOS technology, a high-density crossbar array (CBA) structure is required for memory and logic operations. But one of the critical issues for CBA is the 'sneak path', which corresponds to the parasitic current that bypasses the target storage element and makes the array non-functional[19]. To address this issue, a memory cell in the CBA needs two components: a 'select device' to regulate the current flow and a memory switching element to store the data, which increases both fabrication and operational complexity[20]. Although several reports have shown that it is possible to scale down the switching element to nanometer scale with excellent performance but the challenges arise from cycle-to-cycle (C2C) and device-to-device (D2D) variations due to the stochastic fluctuation and abrupt switching characteristics caused by the overshoot currents during filament formation[21–23]. These variations lead to undesired effects on the adjacent cells in the high-density CBA, thus resulting in breakdown or unintended programming/ erase. Therefore, research studies have been focused to achieve self-compliance (SC) RRAMs that limit current without any external selectors, but achieving array-scale uniformity remains difficult[24–26].

Here, we demonstrate a robust selector-free memory based on a multiferroic $BiFeO_3$ that combines self-compliance with stable bipolar switching, high yield, long endurance, and extended retention. We show that a naturally formed $TiO_x$ interfacial layer at the Ti/$BiFeO_3$ junction renders multiferroic $BiFeO_3$ crossbar memristors intrinsically self-compliant, enabling forming-free bipolar switching without external selector devices while simultaneously delivering high yield, long endurance, and extended retention - resolving the key bottlenecks that have limited the practical deployment of resistive crossbar memories. Beyond previous BFO based NVM devices, our observations include (1) demonstration of BFO-based SCRRAM array with high device yield (~97%) and stable non-volatile resistive switching performance with low C2C and D2D variations; (2) achievement of 4-bit conductance state modulation with statistical evolution; (3) near-ideal noise characteristics with minimal random telegraph noise (RTN) performance, and (4) realization of pulse modulated 4-bit edge computing for digit visualization. These results highlight that the multiferroic BFO has the capability of a scalable platform for oxide-based multibit memory and computing hardware.

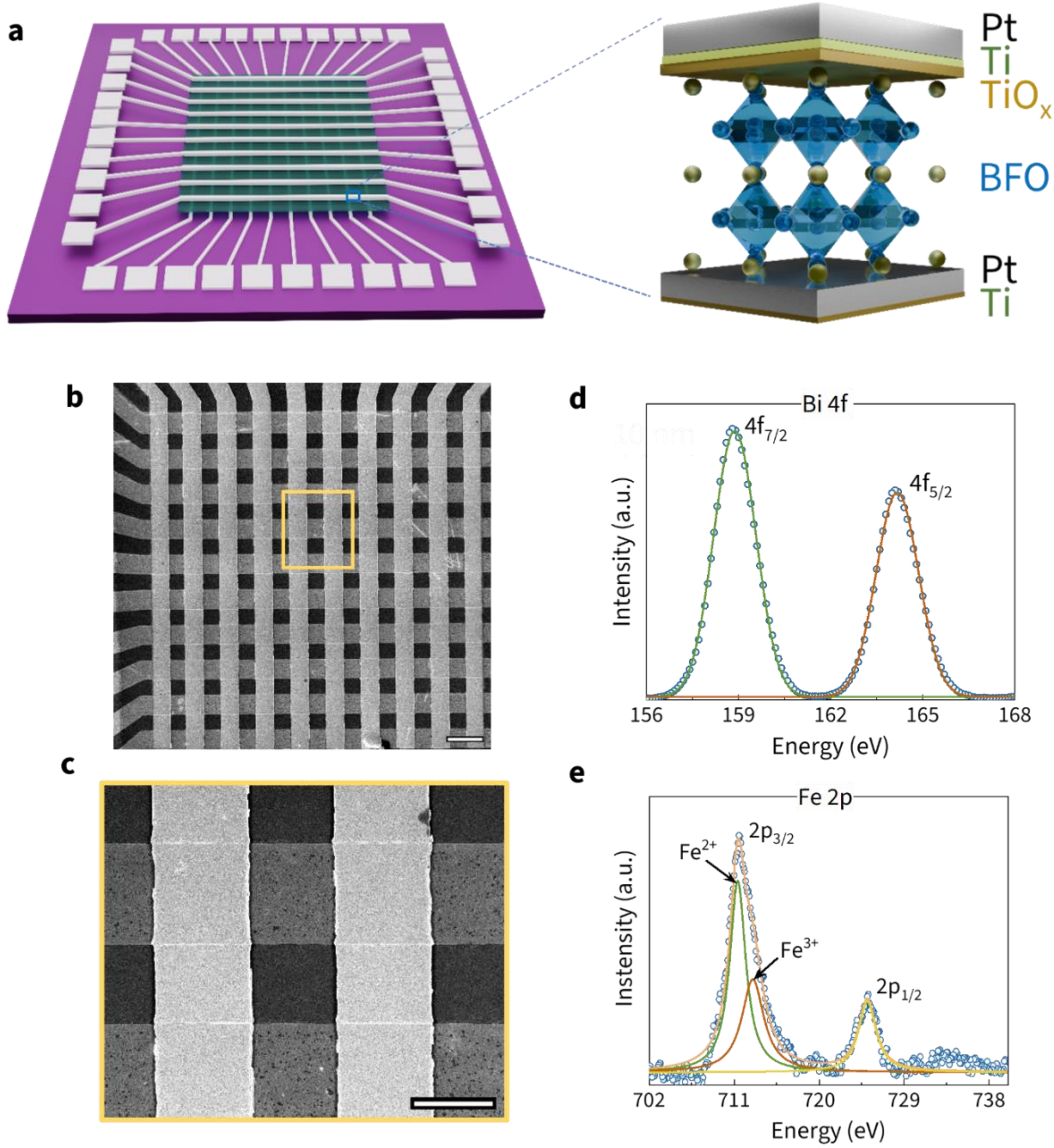

**Fig. 1 | Material characterization and crossbar array fabrication.** ***a,*** *Schematic of the BFO memristive crossbar array. Right panel shows the magnified schematic of a Pt/Ti/$TiO_x$/BFO/Pt/Ti memristor at a certain crosspoint in panel* ***(a)****.* ***b,*** *SEM image of the crossbar array of 10-µm × 10-µm BFO memristor. Scale bar, 20 µm.* ***c,*** *Magnified view of the array at a certain crosspoint in panel* ***(c)****, highlighted in yellow. Scale bar, 10 µm.* ***d-e,*** *XPS measurements showing the core level spectra of Bi 4f* ***(d)****, and Fe 2p* ***(e)*** *of the BFO thin film.*

## Fabrication and characterization

To address the aforementioned challenges, various functional oxide materials have been explored. Among the known ferroelectric oxides, $BiFeO_3$ (BFO) is the most promising and well-studied room-temperature multiferroic material[27]. The discovery of stable multiferroic properties in epitaxial $BiFeO_3$ thin films has marked a major milestone in condensed matter research, enabling a broad range of opportunities for next-generation oxide electronic technologies[28,29]. However, the presence of intrinsic oxygen vacancies in BFO can degrade its ferroelectric performance; intriguingly, these same defects are advantageous for resistive switching in RRAM-based NVM devices[30–32]. Figure 1a represents the schematic of the SC BFO crossbar array architecture, whereas right panel of fig. 1a illustrates the device structure consists of Pt/Ti/$TiO_x$/BFO/Pt/Ti at each crosspoint of the array, where the $TiO_x$ layer is formed naturally during the deposition of the top electrode, reacting with Ti and the surface-absorbed oxygen from the BFO. This $TiO_x$ layer functions as an internal resistor to implement SC, which is necessary to limit the overshoot current during the conducting filament formation in the memristor-based crossbar array without using transistors[20]. The memristor is prepared by directly depositing 28-nm BFO onto a Si/$SiO_2$ substrate with patterned bottom electrodes (50-nm Pt/10-nm Ti), fabricated via e-beam deposition, photolithography, and RF sputtering. The detailed fabrication process is provided in the methods section. Figure 1b-c shows top-view scanning electron microscopy (SEM) images of the fabricated 10 × 10 crossbar array (Fig. 1b) and its magnified 2 × 2 sub-array (Fig. 1c), showing the conformal deposition and smooth surface of the memristor. To investigate the chemical composition, X-ray photoelectron spectroscopy (XPS) is utilized to analyze the elemental oxidation states. The Bi 4f spectrum (Fig. 1d) shows a doublet with peaks at $4f_{7/2}$ and $4f_{5/2}$ consistent with the $Bi^{3+}$ oxidation state typically found in BFO[23]. The Fe 2p spectrum (Fig. 1e) is deconvoluted into contributions from both $Fe^{3+}$ and $Fe^{2+}$ valence states[23]. The presence of $Fe^{2+}$ is frequently observed in BFO thin films and is generally attributed to oxygen vacancies ($V_O$), which act as charge compensators to maintain charge neutrality during the thin film growth process[33]. The O 1s spectrum (Supplementary Fig. 1) typically reveals a primary peak associated with the metal-oxygen (Fe-O/Bi-O) bonds in the perovskite lattice and a secondary high-energy shoulder representing chemisorbed oxygen or oxygen-deficient regions linked to $V_O$ sites[32]. The BFO thin films are further characterized using XRD and Raman scattering as shown in Supplementary Fig. 2.

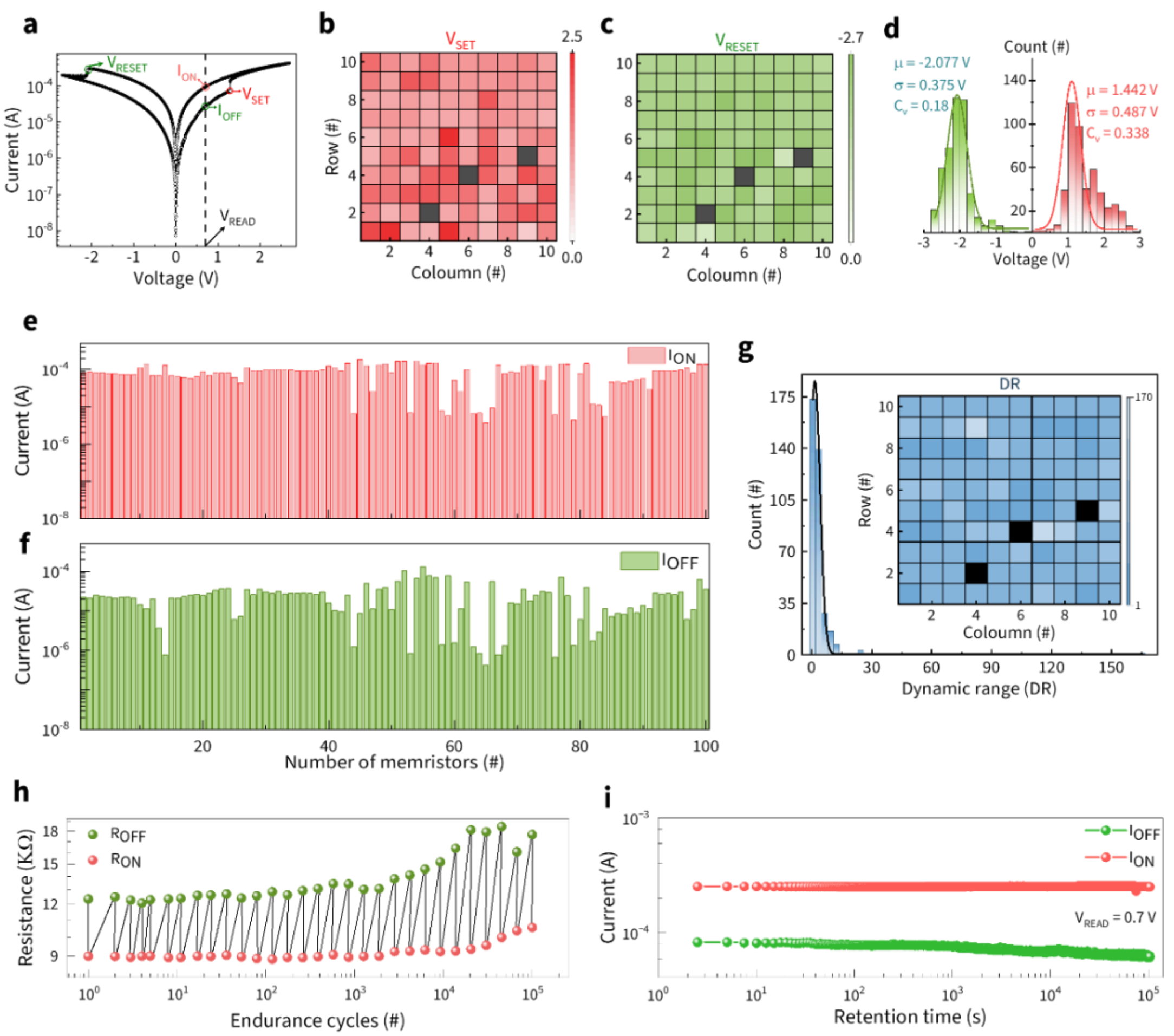

**Fig. 2 | Characterization of the electrical response of the BFO memristive crossbar array.** ***a,*** *A typical representation of I-V characteristics of the BFO crossbar array. At ~ 1.3 V, a sudden jump in current is observed, and this threshold voltage is known as* $V_{SET}$*, which enables the device to switch from the* $I_{OFF}$ *to* $I_{ON}$ *state. In the opposite direction, the current starts to decrease and becomes unstable near ~ -2.0 V. This voltage is known as* $V_{RESET}$*, which facilitates the device to return from* $I_{ON}$ *to* $I_{OFF}$*. The* $I_{ON}$ *and* $I_{OFF}$ *are measured at* $V_{READ}$ *= 0.7 V, of the selected memristor with the top electrode biased and the bottom electrode grounded.* ***b, c*** *The two-dimensional distribution map showing* $V_{SET}$ *(red)* ***(b)*** *and* $V_{RESET}$ *(blue)* ***(c)*** *for each of the memristors (100 memristors) in the 10 × 10 crossbar array. The black colored box within the distribution map indicates the switching failure devices with low DR (<1).* ***d,*** *The cumulative distribution of all the SET and RESET voltage values measured in 100 devices.* ***e-g*** *Statistical histogram of* $I_{ON}$ ***(e)****,* $I_{OFF}$ ***(f)*** *and DR* ***(g)*** *for all the 100 switchable memristors. The*

*inset of **(g)** shows the cumulative distribution of the dynamic range (DR), demonstrating good device uniformity. All switching parameters are extracted at $V_{READ}$ = 0.7 V. **h,** Endurance performance of the memristor for $10^5$ cycles by applying $V_{SET}$ = 2.7 V and $V_{RESET}$ = -3 V for pulse width = 5 μs. **i,** Retention test of the memristor for > $10^5$ s after completing the switching process. $I_{ON}$ and $I_{OFF}$ are measured by applying a bias voltage $V_{READ}$ = 0.7 V.*

**Electrical characteristics**

Figure 2a displays a representative I-V switching curve from a single memristor without requiring a forming process (Supplementary Note 1), capturing the OFF-state current ($I_{OFF}$) and ON-state current ($I_{ON}$) measured at $V_{READ}$ = 0.7 V with SET and RESET voltages around 1.3 V and ~ -2 V, respectively. Each memristor of the 10 × 10 array exhibits non-volatile bipolar resistive switching characteristics similar to those shown in Fig. 2a (Supplementary Fig. 3-6). Notably, all the resistive switching operations are performed without applying current compliance, showing inherent SC characteristics. The presence of SC behavior in the memristor, capable of reducing the overall current level during the switching operations, suppresses the overall energy consumption in the CBA[26]. Additionally, the absence of selector device may require higher set and reset pulse amplitude during the pulse programming scheme, which enhances the possibility of hard breakdown and interference even using V/2 or V/3 bias schemes. Therefore, it's important to incorporate the non-linear characteristics of the device. As shown in Fig. 2a and Supplementary Fig. 3-6, the current changes non linearly with the applied bias voltage, which might be due to the inherent characteristics of BFO, which helps to reduce the current at lower voltage regime. The detailed study of non-linearity behavior for 30 different memristor is shown in Supplementary Fig. 7. The transport mechanism of the Pt/Ti/$TiO_x$/BFO/Pt/Ti memristor indicate a filamentary switching mechanism governed by $V_o$ migration and interfacial barrier modulation for multibit capability (discussed later in the 4-bit conductance state section) driven by the interplay of the applied electric field (Supplementary Fig. 8-9 and Supplementary Note 2).

Figures 2b, c present the statistical distributions of $V_{SET}$ (red) and $V_{RESET}$ (green) across all memristors in the 10 × 10 array, with black square marking switching failures (defined as devices failing to achieve a reasonable dynamic range (DR), i.e., $I_{ON}/I_{OFF} > 1$ at $V_{READ}$), implying a yield of 97%. We statistically analyzed over 400 I-V curves from 100 randomly selected devices (Supplementary Fig. 10) to characterize the variability in switching metrics. To quantify both C2C and D2D variations, we calculated the coefficient of variation ($C_V$), defined as the ratio of the standard deviation (σ) to the mean value (μ), for set and reset voltages in absolute value. The mean value for $V_{SET}$ and $V_{RESET}$ is ~ 1.442 ± 0.487 V and -2.077 ± 0.375 V, respectively, and the corresponding cumulative probability is shown in Supplementary Fig. 11. This relatively low operating window places our device at par with many redox-based oxide[34,35] and 2D-material[36,37] based RRAMs for high-density storage and in-memory computing, where switching often occurs at similar or higher biases depending on the device architecture and filament evolution. The result highlights the effectiveness of the $TiO_x$/BFO heterostructure in enabling low-power bipolar switching while retaining multilevel functionality and stable operation. The minimum C2C variabilities of $V_{SET}$ and $V_{RESET}$ is 13.12% and 6.21%, respectively, and the D2D variabilities rise to 33.8% and 18%, respectively, when considering all 100 devices. As shown in Fig. 2d, all datasets are well described by a normal distribution, indicating good statistical consistency. The observed C2C variation in switching voltage likely stems from the inhomogeneous nature of the filamentary RS, and the D2D variability originates from fabrication-induced structural non-uniformities, including differences in device geometry and active layer thickness[38]. Figure 2e-g shows the statistical histogram of $I_{ON}$, $I_{OFF}$ and DR of all functional memristors in the crossbar array, with the visualization of DR distribution in the inset of Fig. 2g. The average values for $I_{ON}$, $I_{OFF}$ and DR are $8.40 \times 10^{-5}$ A, $2.49 \times 10^{-5}$ A and 4.56, respectively. The cumulative probability distribution of $I_{ON}$ and $I_{OFF}$ is shown in Supplementary Fig. 12. The variation of $I_{OFF}$ compared to $I_{ON}$ is relatively large (in Fig. 2e-f), which could arise due to the incomplete RESET process during the I-V measurements. This observation suggests that both the dynamic range and its uniformity could be further enhanced by achieving a more complete RESET process[35].

The device reliability and stability for practical memory application is further assessed through endurance and retention measurements. To perform these characteristics, a current blind pulse voltage stress is used for endurance cycles, and a current sampling method is used for retention measurements. Figures 2h, i show the endurance and retention performance of a representative device over $10^5$ cycles and $>10^5$ seconds, respectively, exceeding or matching the performance benchmarks of state-of-the-art RRAM memristors and crossbar array architectures (the comparison will be discussed in detail later). Statistical investigation of retention measurements is also performed on randomly selected 50 devices (Supplementary Fig. 13-14), confirming high uniformity across multiple devices. Additionally, the devices are equally robust under continuous I-V sweeps, maintaining functionality beyond 150 consecutive sweeps (Supplementary Fig. 15). The average $I_{ON}$ and $I_{OFF}$ value for retention measurements is $(2.54 \pm 0.11) \times 10^{-4}$ A and $(6.54 \pm 0.27) \times 10^{-5}$ A, respectively, whereas endurance measurements yield (13.86 ± 1.9) KΩ and (9.18 ± 0.41) KΩ (Supplementary Fig. 16). Collectively, these findings highlight that high-yield and reliable switching can be achieved in a standalone configuration without a series access transistor, offering a simplified path towards compact and scalable integration.

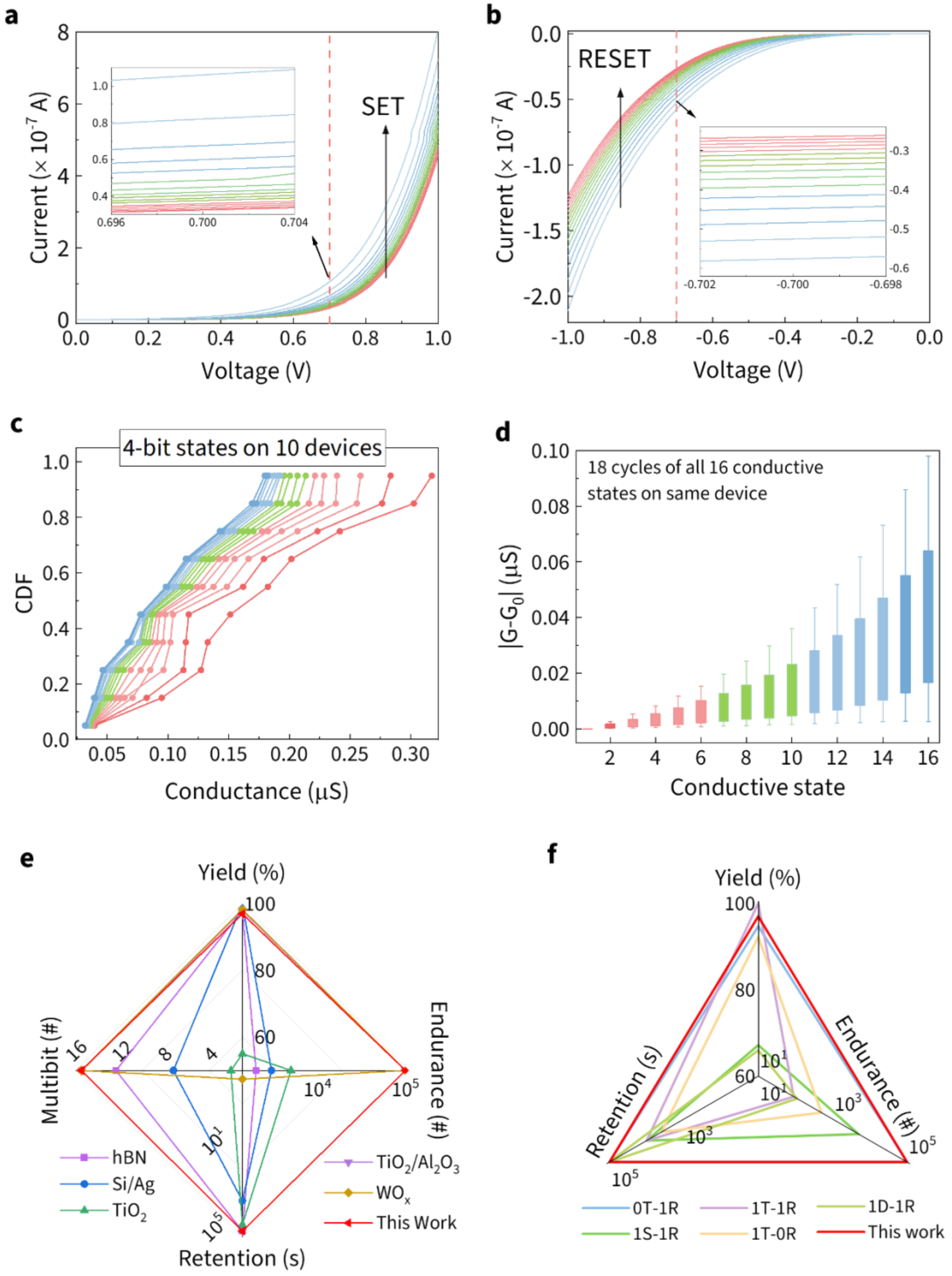

**Fig. 3 | 4-bit switching characteristics.** ***a, b*** *I-V sweep of 16 conductive states for both SET* ***(a)*** *and RESET* ***(b)*** *processes. The inset shows the zoomed version of the low current regime at* $V_{READ}$ *= ± 0.7 V for both the SET and RESET processes, respectively.* ***c,*** *Cumulative distribution function (CDF) plot of all 16 conductive states obtained from 10 different devices*

*using the I–V sweep method during the SET process.* ***d,*** *Statistical distribution of the 4-bit conductance states over 18 switching cycles measured on a single device during the RESET process. The conductance values are normalized to the initial conductance ($G_0$) of each cycle. The boxes indicate the 25$^{th}$ and 75$^{th}$ percentile, while the whiskers denote the minimum and maximum measured values.* ***e,*** *In comparison of $BiFeO_3$ with state-of-the-art oxide and 2D material-based memristors in terms of the yield, endurance cycles, multibit and retention time.* ***f,*** *Spider plot comparing $BiFeO_3$ with five different types of crossbar array architecture, demonstrating the superior performance across all the evaluated metrics.*

**Precise multistate conductance programming**

For neuromorphic computing, memristive devices must offer finely tunable and stable conductance states so that a broad range of artificial intelligence models can be accurately mapped onto the crossbar architecture without adding any additional element like a selector device[10]. Precise control of conductance state is essential not only to avoid the device size and energy consumption associated with using multiple low-precision arrays, but also to enable direct, analog-like implementation of synaptic weight updates for in-memory computing[3,39]. In earlier studies on BFO thin film memristors, multilevel conductance modulation has been demonstrated mostly by adjusting the compliance current during I-V sweeps, and in a limited number of cases by applying sequences of programming voltage pulses. These studies mostly reported on a single device (inconsistency across multiple cycles or across different devices) and generally fell short in terms of programming accuracy, restricting its ability for practical application for high data storage and near/in-memory computing[30–32].

Here, the BFO-based SCRRAM exhibits reproducible 4-bit operation with 16 clearly resolved conductance levels, during both SET and RESET characteristics, as shown in Fig. 3a, b. To demonstrate precise multilevel conductance states, we start with a pristine device by keeping the voltage sweep confined to ± 1 V, below the switching threshold, to suppress abrupt overshoot during the HRS-to-LRS transition. The nearly linear I-V response within ±700 mV and the well-separated read currents at VREAD = 0.7 V (Inset of Fig. 3a, b) underscore the suitability of these devices for precise analog programming. Quantitative assessment of reproducibility is confirmed by cumulative distribution functions constructed from the 16 states in ten independent memristors (Fig. 3c) and by repeatability tests over 18 cycles, where the conductance evolution $|G-G_0|$ remains confined within narrow envelopes during RESET (Fig. 3d). Although high-precision 4-bit programming has been achieved in optimized $HfO_2$- and $TaO_x$-based memoristors[24–26], a comparably systematic, array-scale analysis has been lacking for $BiFeO_3$ and other multiferroic systems; in that context, the state separation and uniformity obtained here place $BiFeO_3$ memristors at par with leading oxide-based and 2D material-based platforms while exploiting the multifunctionality of a room-temperature multiferroic. Alongside the observation of 4-bit conductance states by the voltage sweeping method, we further demonstrate the in-memory computing capability of the BFO memristor by applying precise pulse programming to realize 4-bit edge computing directly within the BFO memristor (discussed in the 4-bit edge computing section).

To evaluate the overall performance of the BFO-based SCRAM, we benchmark against state-of-the-art oxide-based[35,40–42] and 2D material-based[36,37] memristors along with different types of crossbar array architecture[35,43–51]. These comparisons are conducted across several metrics, including yield, retention time, multibit and endurance cycles. Figure 3e indicates that the yield of our BFO device is 97%, which is close to the best reported values and clearly higher than several reported best 2D/oxide systems shown in the comparison. The achievable 4-bit highlights the in-memory computing capability of the devices. It also achieves a continuous endurance cycle of $10^5$ pulses without refreshing, which is one of the most important results for practical memory use. This result significantly outperforms other reported NVM devices and crossbar array architectures. The device also demonstrated stable performance with a retention time of $10^5$ s at room temperature, which suppresses the values in the literature by one order of magnitude. Figure 3f presents a spider plot that offers a detailed comparison of different crossbar array architectures (0T-1R, 1T-1R, 1D-1R, 1S-1R, and 1T-0R) in terms of yield, endurance, and retention time. Our device performance is either better or comparable with other architectures, exhibiting a balanced and superior profile across all metrics. In case of yielding, the 1T-1R shows 100% yielding, though still shorter, retention time and endurance cycles. Only 0T-1R shows a comparable metric for retention and endurance, but falls short for yielding. The benchmarking results, summarized in Fig. 3(e,f), clearly demonstrate that the BFO crossbar device provides high reliability and maintains strong performance across all metrics.

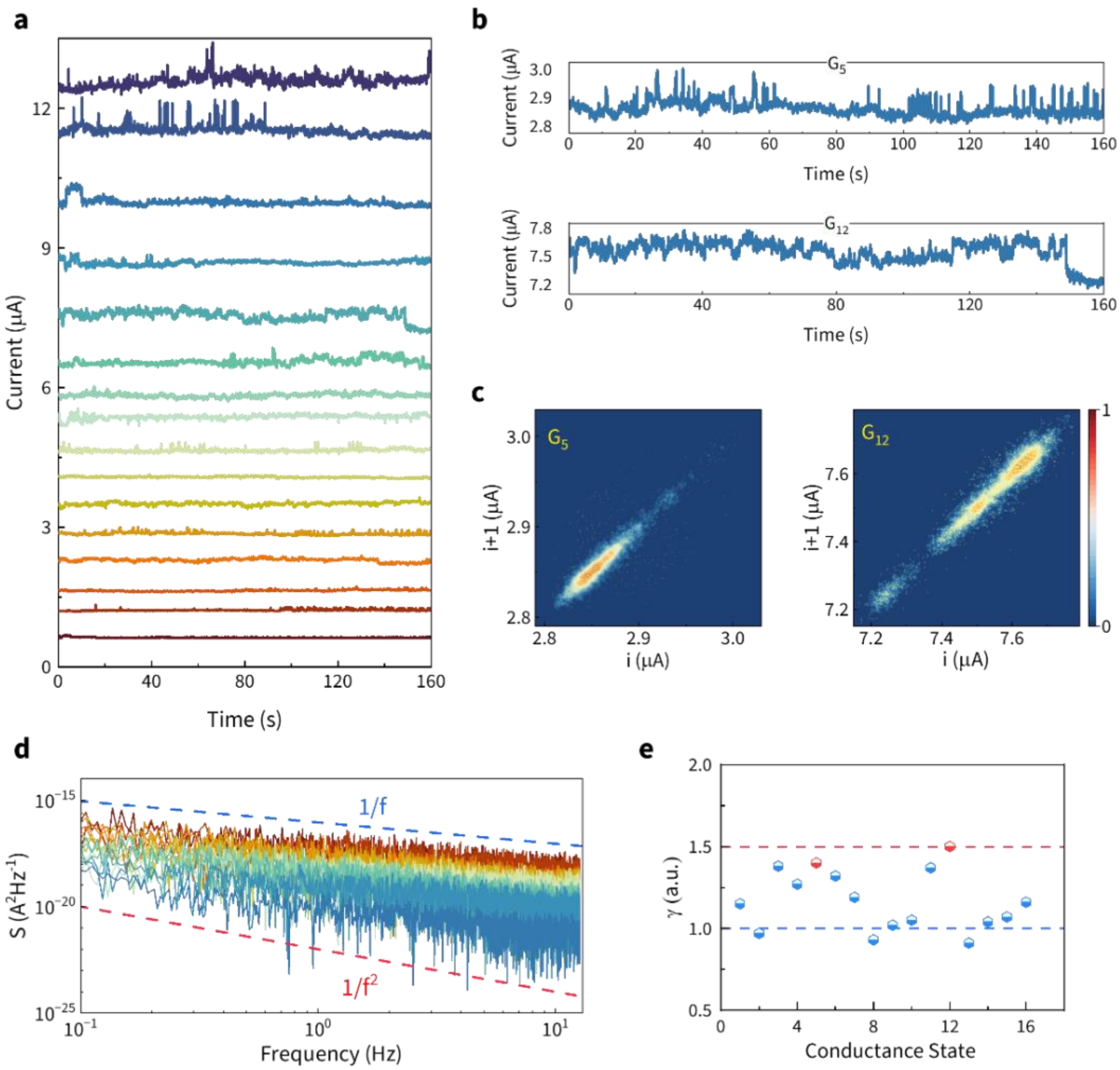

**Fig. 4 | 1/f Noise Characteristics in BFO memristors.** ***a,*** *I-t traces for all 16 conductive states over 160 s.* ***b,*** *Enlarged version of two different I-t traces, revealing the presence of RTN as discrete current step.* ***c,*** *WTLP of the corresponding I-t traces confirm the presence of minimal RTN with most data points clustered into a single group ($G_5$), and $G_{12}$ indicates the presence of a second group/peak.* ***d,*** *Power spectral density (S) versus frequency in log-log representation across all states to reveal near-ideal flicker noise $1/f^{\gamma}$ dependency with $\gamma \approx 1$.* ***e,*** *Spectral exponents ($\gamma$) extracted from $1/f^{\gamma}$ power-law fitting across all accessible conductance states. The two conductive states, $G_5$ and $G_{12}$, in red, correspond to the most*

*noticeable RTN with $\gamma \approx 1.5$, whereas the remaining states show $\gamma \approx 1$, indicating nearly ideal flicker noise behavior.*

**Near-ideal noise characteristics**

Reliable multilevel operation also depends on keeping fluctuations within each conductance level as small as possible, so that adjacent states remain clearly distinguishable. Such conductance fluctuation mainly consists of random telegraph noise (RTN), which is typically characterized by step-like transitions between multiple current levels at random time points under a constant read bias[52]. RTN is a general characteristic of memristive materials and is commonly associated with incomplete or unstable secondary conduction paths that form alongside the main filament[53–55]. Such pathways are highly sensitive to nearby charge-trapping defects and can only be resolved when the measurement sampling rate exceeds the characteristic frequency of the RTN [39], as shown in Supplementary Fig. 17. In time-domain measurements, these instabilities can be effectively suppressed by applying additional "denoising" voltage pulses to stabilize the conductance and preserve usable multilevel states[39]. RTN has been extensively reported in both metal-oxide and two-dimensional memristors, where it's primarily attributed to the charge trapping at point defects located in the vicinity of the conductive filament [53,56,57]. In this context, elucidating RTN behavior in BFO-based memristors is essential for assessing device stability and ensuring reliable multilevel operation for in-memory computing applications.

Here, we report the 1/f noise characteristics of our BFO-based SCRRAM, extracting the frequency scaling of the noise power spectral density across a set of 16 conductance states. Our analysis focuses specifically on RTN, as previous report has shown it to be the primary source of large-amplitude noise in memristive devices, while other noise mechanisms contribute negligibly to overall performance[39]. Time-resolved current measurements are carried out using a source-measure unit (SMU), with the current sampled at a sampling interval of 10 ms. Figure 4a shows the resulting current-time (*I*-*t*) traces for the different conductance levels, in which a voltage of 0.7 V reads each state for 160 s. A magnification of the I-t traces of two conductance states (Fig. 4b) marked as $G_5$ and $G_{12}$, shows the most prominent RTN. To gain a deeper insight into the RTN behavior, the weighted time-lag plot (WTLP) method is used. This method helps to separate discrete current levels, even in the presence of background noise, by analyzing the diagonal of the resulting plots and finding its local maxima; all current levels associated with the RTN signal can be extracted, as each maximum corresponds to a distinct trap-induced current level[37,58]. Figure 4c shows WTLPs of both $G_5$ and $G_{12}$ states by plotting using successive current samples, i + 1 versus i. Additional WTLPs of the remaining 14 states are shown in Supplementary Fig. 18. The WTLP analysis reveals a dominant peak accompanied by a weaker secondary peak, indicating typical RTN behavior with the device current at successive sampling instants is strongly correlated and fluctuating weakly around a single conductance level. This type of RTN signal is correspond to the charge trapping and de-trapping at atomic defects, typically oxygen vacancies or the presence of native defects inside the BFO[53,54,59].

The noise power spectral density for all conductance levels is plotted as a function of frequency in Fig. 4d, together with two dashed reference lines illustrating ideal $1/f^{\gamma}$ behaviour: $\gamma = 1$ (blue, $1/f$) and $\gamma = 2$ (red, $1/f^2$). The low-frequency noise spectra are fitted to extract the frequency exponent ($\gamma$) for each conductance state, and the resulting $\gamma$ values are plotted as a function of conductance state in Fig. 4e. For the majority of the states, we obtain $\gamma \approx 1$, which is characteristic of near-ideal flicker noise rather than white noise or Lorentzian two-level RTN. Only the two conductance states (red symbol) approach $\gamma \approx 1.5$, closer to the ideal two-level RTN signals, i.e., $\gamma \approx 2$, confirming the absence of strong discrete RTN. These measurements demonstrate that our BFO-based SCRRAM maintain stable conductance with predominantly 1/f-type noise across the multilevel operating range, which is favorable for high data storage and near/in-memory computing.

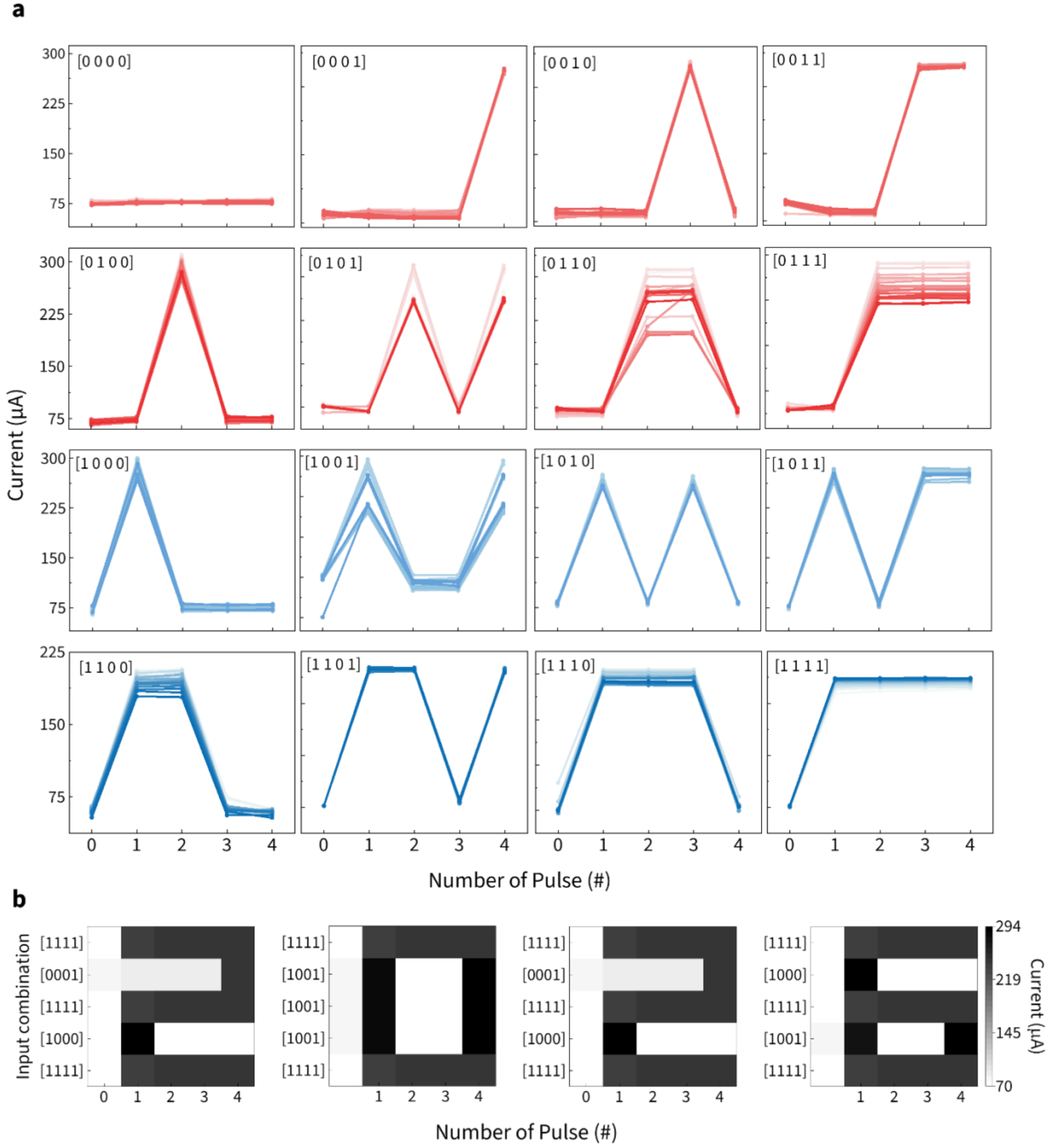

**Fig. 5 | Realization of 4-bit edge computing using the BFO-based SCRRAM.** ***a,*** *20 memristors are randomly selected from a 10 × 10 crossbar array to implement a 4-bit edge computing. For the input "1", a write pulse (-3 V, 1 ms) is used, while for "0", an erase pulse (3 V, 1 ms) is used. The read pulse amplitude and width are 1 V and 1 ms, respectively. The delay between the write/erase and read pulses is 500 μs.* ***b,*** *Digit visualization of "2", "0", "2", "6" by mapping the current response corresponding to the different input conditions '1000' to '1111'.*

**4-bit edge computing**

To evaluate the in-memory computing capability of our BFO memristor, we explore its ability to implement precise pulse-controlled 4-bit edge computing operations directly in the device. Edge computing has emerged as a compelling paradigm for overcoming the latency, energy, and bandwidth limitations associated with conventional cloud-centric architectures, particularly in data-intensive and real-time applications[60]. By enabling data processing directly at or near the source, edge systems minimize data transfer latency while enhancing response speed and energy efficiency[61]. In this context, memristive devices offer a promising hardware platform for edge intelligence, owing to their intrinsic ability to store and process information within the same physical unit[62,63].

To demonstrate the edge computing operations using our BFO-based SCRRAM, we design a programmed pulse scheme consisting of two elementary operations: a write pulse of -3 V for 1 ms to define the "1" state and an erase pulse of +3 V for 1 ms to define the "0" state. Each operation is followed by a read pulse of 0.7 V to measure the device state. These write/erase pulses are arranged into sequences of four input operations, preceded by an initial read step that provides the initial state of the device (Supplementary Fig. 19). In this way, each 4-pulse sequence represents a 4-bit input word $[b_1 b_2 b_3 b_4]$, and the final current level after the operation corresponds to the output of the device for that input. When this pulse protocol is applied to the BFO memristor, the measured current responses (Fig. 5a) distinguish all 16 possible 4-bit input combinations, with each condition repeated on twenty randomly selected devices to confirm reproducibility. This process effectively demonstrates the localized in-memory computing, where the programmed conductance states are directly mapped into measurable current levels without any external computing modules or algorithmic processing. Additionally, to know the latency and the power consumption of the device to achieve these states, we statistically analyze the latency of write/erase state and energy for 10 repeatable states (Supplementary Fig.20). The latency of write/erase is ~ 6 μs, and the power consumption for write and read is ~ 18 nJ and ~ 12.6 nJ, respectively. Such low power consumption underscores the potential of the BFO-based SCRRAM for edge computing applications.

An important application of such systems lies in digit visualization and pattern representation, where analog conductance states can be directly mapped to pixel intensities[63]. In multilevel memristors, distinct and stable current states enable the encoding of grayscale information, allowing visual data to be reconstructed without the need for extensive digital-to-analog conversion. As a practical demonstration, these multilevel currents are mapped to represent numerical digit based on 4-bit logic operations. Figure 5b presents the digits “2”, “0”, “2”, “6” respectively, by reconstructing the current levels with each bit position. This direct mapping facilitates reliable visualization of digit patterns, highlighting the feasibility of in situ image demonstration using intrinsic device physics[64]. The ability to encode and retrieve visual data through controlled modulation of filamentary conduction highlights the potential of BFO memristors for hardware-level neuromorphic vision. Such functionality underscores their suitability for edge computing applications, where compactness, low power operation, and real-time processing are essential.

## Conclusions

In summary, the present work establishes $BiFeO_3$ crossbar arrays with a naturally formed $TiO_x$ interface as a robust selector-free memory platform. By combining self-compliance with stable bipolar switching, high yield, long endurance, and extended retention, the devices address several of the major bottlenecks that have limited practical deployment of resistive arrays. The ability to reproducible 4-bit conductance states, reconfigurable edge computing through pulse programming, together with low 1/f noise signatures across the programmable window, further demonstrates their suitability for multilevel data storage and computing. Our findings push the boundary of $BiFeO_3$ from a well-known room temperature multiferroic oxide to a versatile functional material for high-density memory and near-memory/in-memory computing applications.

## Methodology

### $BiFeO_3$ growth

The BFO active layer is deposited after defining the localized crossbar region through a photolithography process. Following the patterning of the bottom electrode, the exposed active area is utilized for the deposition of the switching layer. BFO thin films are deposited by RF magnetron sputtering using a stoichiometric BFO target. Prior to deposition, the sputtering chamber is evacuated to a base pressure of approximately $1 \times 10^{-6}$ Torr to minimize contamination. The deposition is performed at an RF power of 200 W under a working pressure of 20 mTorr in an Ar atmosphere with a flow rate of 30 sccm. To achieve precise thickness control and ensure uniform film growth, a low deposition rate of ~0.02-0.03 Å $s^{-1}$ are employed. Under these conditions, a deposition time of approximately 3 h is required to obtain a BFO film thickness of ~26 nm. Subsequently, a lift-off process is carried out to define the active BFO region within the crossbar architecture. To enhance crystallinity and optimize the oxygen stoichiometry of the deposited films, post-deposition annealing is performed at 500 °C for 1 h under a low-pressure oxygen atmosphere with an O2 flow rate of 10 sccm. The oxygen annealing treatment facilitated phase stabilization, reduced defect density, and improved the crystalline quality of the BFO layer, thereby providing a suitable active medium for resistive switching applications.

### Device fabrication

The BFO-based self-compliance memristors are fabricated on a 285 nm $SiO_2$/Si substrate. First, an AZ-1505 photoresist is spin coated on the blank substrate and then bottom electrode is patterned using Heidelberg µPG-101 maskless photolithography. Second, 10 nm Ti and on top of that 50 nm Pt is deposited via e-beam evaporation. After deposition, photoresist is removed by immersing the substrate in acetone for 1 hr., thereby completing the fabrication of bottom electrode. The BFO thin film is grown using RF-sputtering method, followed by the additional photoresist patterning to mask the BFO film over the bottom electrode. Finally, similar lithography technique is used following the spin coating of the photoresist to deposit the 50 nm Pt/15 nm Ti top electrode. The BFO based self-compliance memristors are fully fabricated after the lift-off process.

### XPS characterization

X-ray photoelectron spectroscopy (XPS) analysis is performed using a Thermo Scientific™ Nexsa G2™ spectrometer equipped with a micro-focused monochromatic Al $K_\alpha$ source.

Measurements are conducted with a pass energy of 20 eV, and a low-energy electron flood gun is employed to compensate for surface charging effects. An analysis area of 400 × 400 $\mu m^2$ is selected, and spectra are acquired from two different locations to improve reliability, with the final results reported as averaged values. Data analysis is performed using the Thermo Scientific Avantage software package.

**Electrical measurements**

All the electrical characterization are carried out at room temperature using a Lakeshore four probe station integrated with Keithley 4200-SCS semiconductor parameter analyzer. Direct current (DC) measurements are performed using source measure unit (SMU), whereas Keithley 4225-RPM pulse measure unit (PMU) is used for pulse measurements. During all the measurements, the voltage is applied at the top electrode, whereas the bottom electrode is grounded. Random telegraph noise measurements are performed after achieving each conductance state. The current fluctuations are captured for a period of 160 s with a sampling interval of 30 ms and 4096 data points are acquired for each measurement.

**Data availability**

The data that support the findings of this study are available from the corresponding author upon reasonable request.

**Acknowledgements**

S.S., and S.R., thank the Department of Atomic Energy, Government of India for providing financial support. S.S. and P.S. acknowledge Anusandhan National Research Foundation (ANRF) (CRG/2023/006935) for financial support. S.R thank Manas Ranjan Puhan, CSIR-CSMCRI, Bhavnagar for his help in XPS measurements.

**Competing interests**

The authors declare no competing interests.